\documentclass[]{interact} 

\usepackage[english]{babel}
\usepackage[utf8]{inputenc}
\usepackage{amsmath}
\usepackage{amssymb}
\usepackage{type1cm}
\usepackage{bm}
\usepackage{url}
\usepackage{graphicx}
\usepackage{epstopdf}
\usepackage[caption=false]{subfig}

\usepackage[numbers,sort&compress]{natbib}
\bibpunct[, ]{[}{]}{,}{n}{,}{,}

\theoremstyle{plain}

\theoremstyle{definition}

\theoremstyle{remark}

\begin{document}

\title{
Variable screening using factor analysis for high-dimensional data with multicollinearity
}

\author{
    \name{
    Shuntaro Tanaka\textsuperscript{a, b}
    \thanks{CONTACT Shuntaro Tanaka. Email: tanaka.shuntaro@jri.co.jp} 
    and 
    Hidetoshi Matsui\textsuperscript{c}
    \thanks{CONTACT Hidetoshi Matsui. Email: hmatsui@biwako.shiga-u.ac.jp} 
    }
    \affil{
        \textsuperscript{a}The Japan Research Institute,Ltd., Japan; 
        \textsuperscript{b}Graduate School of Data Science, Shiga University 1-1-1 Banba Hikone Shiga 522-8522 Japan;
        \textsuperscript{c}Faculty of Data Science, Shiga University 1-1-1 Banba Hikone Shiga 522-8522 Japan;
    }
}

\maketitle

\begin{abstract} 
\noindent
Screening methods are useful tools for variable selection in regression analysis when the number of predictors is much larger than the sample size.
Factor analysis is used to eliminate multicollinearity among predictors, which improves the variable selection performance. 
We propose a new method, called Truncated Preconditioned Profiled Independence Screening (TPPIS), that better selects the number of factors to eliminate multicollinearity. 
The proposed method improves the variable selection performance by truncating unnecessary parts from the information obtained by factor analysis.
We confirmed the superior performance of the proposed method in variable selection through analysis using simulation data and real datasets.
\end{abstract}


\section{Introduction}
Recent developments in the field of communication technology have generated data in a variety of fields, including finance, medicine, and agriculture. 
Appropriate analysis of such data enables us to reveal the relationships inherent in the complex phenomena.
Regression analysis is one of the most widely used statistical methods to do this.
For example, if we want to understand the regularity of the sales of a product, we set the sales as the response and the product attributes (price, color, size, etc.) as the predictors.
To understand the correct relationship between the predictors and the response, it is necessary to select and analyze important variables from the large amount of data that appear to be strongly related to a given response.

Variable selection is used in several fields. 
In finance, variables related to corporate accounting data are selected to construct a statistical model that predicts the risk of corporate bankruptcy \cite{tian2015variable}. 
Another example is the selection of variables that relate to data on macroeconomic indicators to estimate volatility, which is used to select which company to invest in and to make decisions about the timing of investments \cite{fang2020predicting}. 
Variable selection is also used in clinical models that predict possible future diseases \cite{chowdhury2020variable} and in near-infrared spectroscopy analysis to measure food compositions \cite{yun2019overview}.

It is difficult to apply the classical variable selection techniques such as stepwise regression to high-dimensional data. 
Methods using $L_1$-type regularization also fail to select variables for ultra-high dimensional data. 
While more recently, Sure Independence Screening (SIS) was proposed to greatly reduce the dimension of the predictors and select important variables \cite{fan2008sure}.  
SIS selects predictors in the order of their Pearson's correlations with the response in linear regression models.
Although this is a simple technique, the probability that the set of variables selected by SIS contains a set of truly important variables converges to 1 as the sample size increases. 
Several extensions of SIS have been proposed. 
\cite{fan2010sure} extended the idea of SIS to generalized linear models, and 
\cite{fan2011nonparametric} extended it to high-dimensional additive models. 
In addition, there are screening methods that use non-linear correlations instead of Pearson correlations.
\cite{li2012robust} proposed a method that is robust to outliers that uses Kendall's rank correlation coefficient.
\cite{li2012feature} used distance correlation, and 
\cite{balasubramanian2013ultrahigh} used the Hilbert-Schmidt Independence Criterion (HSIC).
With these criteria, we can apply the screening methods without assuming any distribution for the variables.
\cite{zhang2017correlation} also proposed a method for censored data. 
The development of screening methods was summarized in
\cite{fan2018sure}.

However, most of these screening methods have the problem that their performance degrades in the presence of multicollinearity. 
To solve this problem, \cite{wang2016high} proposed a method called High-dimensional Ordinary Least squares Projection (HOLP), which accommodates highly multicollinear predictors by selecting variables in the order of their relations estimated by high-dimensional ordinary least squares. 
Factor Profiled Sure Independence Screening (FPSIS) proposed by 
\cite{wang2012factor} transforms the data for predictors by applying factor analysis, which reduces multicollinearity. 
Then we can select appropriate variables by applying SIS to the transformed data that correspond to unique factors.
Preconditioned Profiled Independence Screening (PPIS) proposed by 
\cite{zhao2020high} improved the FPSIS transformation process to better reduce multicollinearity. 
PPIS eliminates unnecessary information from the predictors by using all of the common factors obtained from applying factor analysis to the predictors, whereas FPSIS uses only a subset of common factors.

However, PPIS seems to eliminate more information about predictors than necessary, which can degrade variable selection performance.
To overcome this issue, we propose a method to improve the effectiveness of removing multicollinearity by modifying PPIS to select variables more accurately.
We truncate some of the common factors eliminated in the PPIS transformation process to prevent excessive loss of information for variable screening.
We call our proposed method Truncated PPIS (TPPIS). 
The reason why TPPIS improves the variable selection performance can be explained by a model based on the distribution of eigenvalues.
The truncation part is determined objectively using the BIC-type criterion proposed by 
\cite{wang2012factor}. 
SIS is then applied to the data whose multicollinearity has been removed by the transformation process.
Through analysis of simulated and real data, we show that TPPIS can transform data appropriately.

The remainder of this paper is organized as follows.
Section 2 describes existing screening methods, and then the proposed method is described in Section 3.
In Section 4, we confirm the performance of the screening method through a simulated data analysis, and then report the results of real data analysis in Section 5.
Section 6 summarizes the main points.

\section{Screening methods utilizing factor analysis}



Suppose we have $n$ sets of observations $\{ (y_{i}, x_{i}), i=1,\ldots,n \}$, where $y_{i} \in \mathbb{R}$ is a response and $\bm{x}_{i} = (x_{i1}, \ldots, x_{ip})^{T} \in \mathbb{R}^{p}$ is a vector of predictors.
In particular, we assume that $n<p$ and $\bm{x}_{i}$ is standardized and $y_{i}$ is centered. 
The relationship between $y_{i}$ and $\bm{x}_{i}$ is assumed to be represented by the following linear model.
\begin{equation}
y_{i} = \bm{x}_{i}^{T} \bm{\beta} + \varepsilon_{i},
\nonumber
\end{equation}
where $\bm{\beta} = (\beta_{1}, \ldots, \beta_{p})^{T} \in \mathbb{R}^{p}$ are regression coefficients and $\varepsilon_{i} \in \mathbb{R}$ is independent and identically distributed (i.i.d.) random noise following $N(0, \sigma^{2})$.
Let $\bm{y} = (y_{1}, \ldots, y_{n})^{T} \in \mathbb{R}^{n}$, $X = (\bm{x}_{1}, \ldots, \bm{x}_{n})^{T} \in \mathbb{R}^{n \times p}$, and $\bm{\varepsilon} = (\varepsilon_{1}, \ldots, \varepsilon_{n})^{T} \in \mathbb{R}^{n}$. 
Then the above linear model can be expressed as
\begin{equation}
    \bm{y} = X \bm{\beta} + \bm{\varepsilon} \label{linear}.
\end{equation}

Let $\bm{\omega} = ( \omega_{1}, \ldots, \omega_{p} )^{T} = {X}^{T} \bm{y} \in \mathbb{R}^{p}$ and define the importance of the $j$-th variable as $|\omega_{j}|$ ($1 \leq j \leq p$). 
SIS excludes predictors that are considered to be unnecessary by selecting the $j$-th variables in order of increasing $|\omega_{j}|$.
However, SIS does not work well in the presence of strong multicollinearity.
For example, $|\omega_{j}|$ becomes smaller even for important variables or $|\omega_{j}|$ becomes larger even for unimportant variables.

In FPSIS \cite{wang2012factor}, SIS is applied after a transformation process to remove multicollinearity by applying factor analysis. 
Let $Z \in \mathbb{R}^{n \times d}$ be a matrix of vectors of $d$ $(<n)$ common factors of $X$, $B \in \mathbb{R}^{p \times d}$ be factor loadings, and $\check{X} \in \mathbb{R}^{n \times p}$ be a matrix composed of unique factors.
Then we can express their relationships as $X = Z B^{T} + \check{X}$, where the columns of $\check{X}$ are independent each other. 
Although $Z$ is not uniquely determined due to the rotation invariance, a solution for $Z$ can be obtained by singular value decomposition.

Let $\mu_{1}, \ldots, \mu_{n}$ be $n$ singular values of $X$, where $\mu_{1} \geq \ldots \geq \mu_{n} > 0$, since we assume $n<p$ here.
The singular value decomposition of $X$ gives
\begin{equation}
    X = U D V ^{T}, 
    \label{svd}
\end{equation}
where $U = (\bm{u}_{1}, \ldots, \bm{u}_{n}) \in \mathbb{R}^{n \times n},
\bm{u}_{l} = (u_{1l}, \ldots, u_{nl})^{T} \in \mathbb{R}^{n},
D = \text{diag}(\mu_{1}, \ldots, \mu_{n}) \in \mathbb{R}^{n \times n},
V = (\bm{v}_{1}, \ldots, \bm{v}_{n}) \in \mathbb{R}^{p \times n},
\bm{v}_{l} = (v_{1l}, \ldots, v_{pl})^{T} \in \mathbb{R}^{p}$ $(l=1,\ldots,n)$,
and 
$U^{T} U = V^{T} V = {I_{n}}$.
Let $U_{1} = (\bm{u}_{1}, \ldots , \bm{u}_{d}) \in \mathbb{R}^{n \times d}$ denote the first $d$ columns of the matrix $U$ in \eqref{svd}. 
Then $U_{1}$ can be regarded as one of the solutions of $Z$.
\cite{wang2012factor} decided the value of $d$ by the following equation using the ratio of the singular values of $X$: 
\begin{equation}
\label{d}
d = \underset{1 \leq l \leq n-1} {\operatorname{argmax}} \frac{\mu_{l}^{2}}{\mu_{l+1}^{2}}.
\end{equation}
The projection matrix onto the orthogonal complement of the linear subspace spanned by the column vectors of the matrix $U_{1}$ is given by 
\begin{equation}
Q_{F} = I_{n} - U_{1} ( U_{1}^{T} U_{1} )^{-1} U_{1}^{T}.
\label{Q_F}
\end{equation}
Left-multiplying both sides of \eqref{linear} by $Q_{F}$ gives
\begin{equation}
  Q_{F} \bm{y} = Q_{F} X \bm{\beta} + Q_{F} \bm{\varepsilon}. \label{linear_q}
\end{equation}
\noindent
Let $\hat{\bm{y}} = (\hat{y}_{1}, \ldots,  \hat{y}_{n})^{T}= Q_{F} \bm{y}$ and $\hat{X} = (\hat{\bm{x}}_{1}, \ldots , \hat{\bm{x}}_{n})=Q_{F}X$. 
$\hat{X}$ is an approximation of the unique factors $\check{X}$.
The use of $\hat X$ instead of $X$ enables us to eliminate multicollinearity and to select appropriate variables.
FPSIS calculates $\bm{\omega} = (\omega_{1}, \ldots, \omega_{p})^{T} 
= \hat{X}^{T} \hat{\bm{y}} \in \mathbb{R}^{p}$, and then selects variables where $|\omega_{j}|$ is large in order.  

PPIS \cite{zhao2020high} improved the FPSIS transformation process.
First, after applying SVD to $X$ as in \eqref{svd}, they divided each of the matrices $U,D,V$ into two parts at the $d$-th column: 
$U_{1} = (\bm{u}_{1}, \ldots , \bm{u}_{d}) \in \mathbb{R}^{n \times d}$, 
$U_{2} = (\bm{u}_{d+1}, \ldots \bm{u}_{n}) \in \mathbb{R}^{n \times (n-d)}$, 
$D_{1} = \text{diag}(\mu_{1}, \ldots , \mu_{d}) \in \mathbb{R}^{d \times d}$, 
$D_{2} = \text{diag}(\mu_{d+1}, \ldots , \mu_{n}) \in \mathbb{R}^{(n-d) \times (n-d)}$,
$V_{1} = (\bm{v}_{1}, \ldots , \bm{v}_{d}) \in \mathbb{R}^{p \times d}$, 
$V_{2} = (\bm{v}_{d+1}, \ldots \bm{v}_{n}) \in \mathbb{R}^{p \times (n-d)}$.
Let 
\begin{equation}
Q_{P} = U_{2}D_{2}U_{2}^{T} \left\{ I_{n} - U_{1} ( U_{1}^{T} U_{1} ) U_{1}^{T} \right\}
\label{Q_P}
\end{equation}
and replace $Q_{F}$ with $Q_{P}$ in \eqref{linear_q}.
This is based on the Puffer transformation \cite{jia2012preconditioning}.
PPIS calculates $\bm{\omega} = \hat{X}^{T} \hat{\bm{y}}$ as in FPSIS, where $\hat{\bm y} = Q_P\bm y$ $\hat X = Q_PX$, and then selects variables in order of the size of $|\omega_{j}|$. 
The number of dimensions $d$ of $U_{1}$ is determined by \eqref{d} using the ratio of the singular values of $X$.
We explain the reasonableness of PPIS in Section \ref{Reasons why TPPIS improves the effectiveness of removing multicollinearity} using a model based on the distribution of eigenvalues.

However, if the magnitudes of the singular values after the $d$-th are not sufficiently small compared to those before the $d$-th, $\hat{X}$ is still multicollinear when we simply remove from $X$ the effects that are related to the first $d$ common factors of $X$.
Therefore, by removing the influence of the $n$ common factors of $X$ including the information after the $d$-th factor that is not used in FPSIS, $\hat{X}$ becomes closer to the unique factors $\check{X}$, which leads to the elimination of more multicollinearity.

\section{Proposed method}


\subsection{TPPIS}
\label{TPPIS}
We propose selecting the number of factors to eliminate more multicollinearity by modifying the transformation process in PPIS. 
Let $\alpha$ be a tuning parameter that satisfies $\alpha \in (0,1]$ and $d<[n \alpha]$.  
After applying SVD to $X$, as in \eqref{svd}, we divide $U,D,V$ into three parts at the $d$-th column and the $[n\alpha]$-th column: 
$U = (U_{1}, U_{2a}, U_{2b})$,
$D = \text{diag} (\mu_{1}, \ldots, \mu_{n})$,
$V = (V_{1}, V_{2a}, V_{2b})$, 
$U_{1} = (\bm{u}_{1}, \ldots, \bm{u}_{d})$, 
$U_{2a} = (\bm{u}_{d+1}, \ldots, \bm{u}_{[n\alpha]})$, 
$U_{2b} = (\bm{u}_{[n\alpha]+1}, \ldots, \bm{u}_{n})$, 
$D_{1} = \text{diag}(\mu_{1}, \ldots, \mu_{d})$, 
$D_{2a} = \text{diag}(\mu_{d+1}, \ldots, \mu_{[n\alpha]})$, 
$D_{2b} = \text{diag}(\mu_{[n\alpha]+1}, \ldots, \mu_{n})$, 
$V_{1} = (\bm{v}_{1}, \ldots, \bm{v}_{d})$, 
$V_{2a} = (\bm{v}_{d+1}, \ldots, \bm{v}_{[n\alpha]})$, and
$V_{2b} = (\bm{v}_{[n\alpha]+1}, \ldots, \bm{v}_{n})$.
Then we define the following projection matrix
\begin{equation}
Q_{T} = U_{2a} D_{2a}^{-1} U_{2a}^{T} \left\{ I_{n} - U_{1} ( U_{1}^{T} U_{1} ) U_{1}^{T} \right\}.  
\label{Q_T}
\end{equation}
Using $\hat X = Q_T X$ rather than $Q_P X$, we can eliminate multicollinearity more accurately since $Q_{T}$ leaves the information that corresponds to the unique factors by truncating $U_{2b}$ and $D_{2b}$ from $U_{2}$ and $D_{2}$, respectively. 
TPPIS calculates $\hat{\bm{y}}, \hat{X},$ and $\bm{\omega}$ using the equation that replaces $Q_{F}$ with $Q_{T}$ in \eqref{linear_q}, and then selects variables where $|\omega_{j}|$ is large in order.

Denote a set of $k$ selected variables as 
\begin{equation}
M_{k} = \{ 1 \leq j \leq p : |\omega_{j}| \text{ is among the first } k \text{ largest of all } \}
\nonumber
\end{equation}
and denote predictors whose columns are composed of $M_{k}$ as $X(M_{k})\in \mathbb{R}^{n \times k}$.
We predict the response using $\bm{y} = X(M_{k}) \hat{\bm{\beta}}(M_{k})$, where $\hat{\bm{\beta}}(M_{k})$ is the least squares estimator of the regression coefficient of $\hat{X}(M_{k})$, that is, 
\begin{equation}
\hat{\bm{\beta}}(M_{k})
= \left\{ \hat{X}(M_{k})^{T} \hat{X}(M_{k}) \right\} ^{-1} \hat{X}(M_{k})^{T} \hat{\bm{y}}.
\label{beta_M_k}
\end{equation}

\subsection{Reasons why TPPIS improves the effectiveness of removing multicollinearity}
\label{Reasons why TPPIS improves the effectiveness of removing multicollinearity}
We discuss the reason why TPPIS improves the effectiveness of removing multicollinearity and the variable selection performance. 
\cite{zhao2020high} indicates that the transformation process using $Q_{P}$ of \eqref{Q_P} works well for data that follow a highly multicollinear spike model. 
The spike model has the property that some eigenvalues of the variance-covariance matrix are larger than others. 
Suppose that the eigenvalues of a variance-covariance matrix $X$, denoted by $\Sigma_{p}$, can be divided into three size categories: large, medium, and small.
Among $p$ eigenvalues, let $d$ be the number of large eigenvalues, $m$ be the number of medium eigenvalues, and $p-d-m$ be the number of small eigenvalues.
Then the spike model assumes that 
$\Sigma_{p}$ is represented as 
\begin{equation}
\Sigma_{p}
=\sum_{r=1}^{d} (\lambda_{r}+\sigma_{0}^{2}) \bm{u}^{*}_{r} {\bm{u}^{*}_{r}}^{T}
+\sum_{s=1}^{m} (\omega_{s}+\sigma_{0}^{2}) \bm{u}^{*}_{d+s} {\bm{u}^{*}}^{T}_{d+s}
+\sum_{t=1}^{p-d-m} \sigma_{0}^{2} \bm{u}^{*}_{d+m+t} {\bm{u}^{*}}^{T}_{d+m+t}
,
 \nonumber
\end{equation}
where $\lambda_{1} \geq \ldots \geq \lambda_{d} > \omega_{1} \geq \ldots \geq \omega_{m} > 0$, $\sigma_{0}^{2}$ is a positive constant, and $\{ \bm{u}^{*}_{1}, \ldots,\bm{u}^{*}_{p} \}$ constitute an orthonormal basis of $\mathbb{R}^{p}$. 
In this case, $X$ can be expressed as
\begin{equation}
  X = \sum^{d}_{r=1} \sqrt{\lambda_{r}} \bm{z}_{r} {\bm{u}^{*}_{r}}^{T}
    + \sum^{m}_{s=1} \sqrt{\omega_{s}} \bm{z}_{d+s} {\bm{u}^{*}_{d+s}}^{T}
    + \sigma_{0}^{2} \Lambda
    , 
      \label{spiked_model}
\end{equation}
where $\bm{z}_{w} \in \mathbb{R}^{n}$ $(w=1,\ldots,d+m)$ are i.i.d. $N({\bf 0}, I_{n})$ vectors and $\Lambda \in \mathbb{R}^{n \times p}$ has i.i.d. $N(0, 1)$ elements.
The vectors $\bm{z}_{r}$ and $\bm{u}^{*}_{r}$ respectively represent a common factor and a factor loading of $X$, and $\sigma_{0}^{2} \Lambda$ represents a unique factor of $X$. 
Let $X_{1}, X_{2}, X_{3}$ be the first, second, and third terms of \eqref{spiked_model}, respectively; that is, we can express \eqref{spiked_model} as $X = X_{1} + X_{2} + X_{3}$.

Since $Q_{F}$ in \eqref{Q_F} is the projection matrix onto the orthogonal complement of the linear subspace spanned by the column vector $U_{1} \in \mathbb{R}^{n \times d}$, $Q_{F}$ can remove the effect of $d$ common factors.
That is, 
\begin{equation}
\begin{aligned} 
  Q_{F} X &= Q_{F} (X_{1} + X_{2} + X_{3})\\
          &\approx X_{2} + X_{3} .
  \nonumber
\end{aligned} 
\end{equation}
The PPIS transformation process using $Q_{P}$ in \eqref{Q_P} can remove the effect of $X_{1}$ and $X_{2}$. 
However, since $U_{2}$ and $D_{2}$ in $Q_{P}$ use all column vectors after the $d$-th column, some extra information seems to have been removed from the unique factor $X_{3}$ that should have been left behind. 
$Q_{T}$ in \eqref{Q_T}, which truncates $U_{2b}$ from $U_{2}$ and $D_{2b}$ from $D_{2}$, can improve variable selection performance by leaving the unique factors more accurately.

\subsection{Selection of tuning parameter}
The performance of the proposed method strongly depends on the dimension $d$ of $U_{1}$, the tuning parameter $\alpha$, and the number $k$ of selected variables.
We have to decide appropriate values for them.
To do this, we use the BIC-type criterion adapted to high-dimensional data proposed by \cite{wang2012factor}.
Using $\hat{\bm{\beta}}(M_{k})$ in \eqref{beta_M_k}, the BIC-type criterion is given by
\begin{equation}
\begin{aligned} 
    \text{BIC}(M_{k}) = \text{log} \left\{ \left|\left| \bm{y} - X(M_{k}) \hat{\bm{\beta}}(M_{k}) \right|\right| ^{2} \right\} + (n^{-1} \text{log }p) |M_{k}| \text{ log }n.
    \label{bic}
\end{aligned}
\end{equation}
We use grid search to find the optimal $d$, $\alpha$, and $k$, 
selecting the values with which make BIC smallest as the optimal parameters.

\section{Simulation examples}
\label{Simulation examples}
To investigate the effectiveness of the proposed TPPIS method, we compare TPPIS with the existing methods. 
After calculating the importance of each predictor on the response for each method, the number of variables is determined using the BIC-type criterion \eqref{bic}, and then the variable selection performance is verified.

\subsection{Settings for simulated data}
\label{Settings for simulated data}
We conducted four examples. 
The sample size $n$ and the number of predictors $p$ are set as $n= 100, 300$, and $p = 1000$ as common values for each example, respectively. 
For the TPPIS parameter $d$, we examined six patterns: $0.2n, 0.4n, 0.6n, 0.8n, 1.0n$, and the value given by \eqref{d}.
In addition, we examined five values ranging from 0.2 to 1.0 in increments of 0.2 for $\alpha$.
For the number of variables, $k$, we examined $p$ values ranging from 1 to $p$.
We then select the $d$, $\alpha$, and $k$ giving the smallest BIC as the optimal parameters.

\begin{itemize}
\item Example 1

For each $i$ in $1 \leq i \leq n$, 
\begin{equation}
y_{i} = 5 x_{i1} + 5 x_{i2} + 5 x_{i3} - 15 x_{i4} + \varepsilon_{i}, 
\nonumber
\end{equation}
where $\varepsilon_{i}$ are i.i.d.\ errors following $N(0,1)$, $\bm{x}_{i}=(x_{i1}, \ldots , x_{ip})^{T}$ are i.i.d.\ predictors following $N({\bf 0}, \Sigma)$ and the variance-covariance matrix $\Sigma=(\Sigma_{jk})^{p}_{j,k=1}$ satisfies 
\begin{equation}
\begin{aligned} 
\Sigma_{jj} &= 1, \\
\Sigma_{jk} &= \varphi\ (j \neq k, j \neq 4, k \neq 4), \\
\Sigma_{4,k} &= \Sigma_{j,4}= \sqrt{\varphi}\ (j, k \neq 4)
.
\nonumber
\end{aligned} 
\end{equation}
We investigated three values for the parameter $\varphi$: 0.5, 0.7, and 0.9.\\

\item Example 2

For each $i$ in $1 \leq i \leq n$, 
\begin{equation}
y_{i} = 5 x_{i1} + 5 x_{i2} + 5 x_{i3} - 15 x_{i4} + 5 x_{i5} + \varepsilon_{i} . \nonumber
\end{equation}
The setting is similar to that in Example 1, but the fifth variable is added.
In addition, the variance-covariance matrix $\Sigma$ of the predictor satisfies $\Sigma_{5,j}=\Sigma_{j,5}= 0$ $(j\neq 5)$.\\

\item Example 3

For each $i$ in $1 \leq i \leq n$, 
\begin{equation}
y_{i} = 5 x_{i1} + 5 x_{i2} + 5 x_{i3} - 15 x_{i4} + 5 x_{i5} + \varepsilon_{i} . \nonumber
\end{equation}
The regression model is the same as in Example 2, except that the sixth variable, which is not included in the regression model, satisfies $x_{i6} = 0.8 x_{i5} + \delta_{i}$, where $\delta_{i}$ follows i.i.d. $N(0, 0.01)$.
Compared to Example 2, the data for the predictors are more multicollinear.\\

\item Example 4

We consider the case where $X$ follows a spike model \eqref{spiked_model}, given by 
\begin{equation}
  X = \sum^{d}_{r=1} \bm{z}_{r} \bm{b}_{r}^{T} 
      + \sum^{m}_{s=1} n^{\frac{-(s+9)}{m+10}} \bm{z}_{d+s} \bm{b}_{d+s}^{T}
      + \check{X} 
      \nonumber
      ,
\end{equation}
where $\bm{z}_{k} \in \mathbb{R}^{n}$ $(k=1, \ldots, d+m)$ are i.i.d.\ vectors following $N({\bf 0}, I_{n})$, $\bm{b}_{k} \in \mathbb{R}^{p}$ is a vector of i.i.d. $N(0, 1)$ elements, and $\check{X} = (\bm{\check{x}}_{1}, \ldots, \bm{\check{x}}_{n})^{T} \in \mathbb{R}^{n \times p}$ with $\bm{\check{x}}_{i} = (\check{x}_{i1}, \ldots, \check{x}_{ip})^{T} \in \mathbb{R}^{p}$, $E(\check{x}_{ij})=0$, and $\text{cov}(\check{x}_{ij_{1}}, \check{x}_{ij_{2}})=I_{p}$.
This case corresponds to equation \eqref{spiked_model} with $\sqrt{\lambda_{r}} = 1 \ (1 \leq r \leq d), \sqrt{\omega_{s}}  = n^{\frac{-(s+9)}{m+10}} \ (1 \leq s \leq m),$ and $ \sigma_{0}^{2} = 1 $.
This model is the same as that used in the simulation by \cite{zhao2020high}.

In this example, $d$ is set to 3 and $m$ is set according to 4 patterns: $0.2n$, $0.4n$, $0.6n$, and $0.8n$. 
The regression model is given by 
\begin{equation}
y_{i} = 5 x_{i1} + 4 x_{i2} + 3 x_{i3} + 2 x_{i4} + \varepsilon_{i}, 
\nonumber
\end{equation}
where
$\varepsilon_{i}$ are i.i.d.\ errors following $ N(0, \sigma^{2})$ with $\sigma^{2} = $ var$(X \bm{\beta}) / 5$ and $\bm{\beta}=(5,4,3,2,0,\ldots,0)^{T} \in \mathbb{R}^{p}$.

\end{itemize}

In each example, we generate datasets 100 times for each combination of parameters.
For each dataset, the numbers of selected predictors and the least squares estimator \eqref{beta_M_k} is calculated.
The number of variables is determined using the BIC in \eqref{bic}.

\subsection{Comparison methods}
The proposed TPPIS method is compared with the existing SIS, FPSIS, and PPIS methods.
In addition to the original FPSIS which selects the value of $d$ using the ratio of eigenvalues \eqref{d}, we also compared a modified FPSIS where $d$ is selected by the BIC in \eqref{bic} rather than \eqref{d}. 
We denote this method as FPSIS$_{BIC}$.
We test the values of $d$ in FPSIS$_{BIC}$ with six patterns, as in the case of TPPIS.

\subsection{Score metric for screening}
We evaluate the variable selection performance of the screening methods using the score based on the number of correctly and incorrectly selected variables.
We refer to necessary predictors as Positive (P) and unnecessary variables as Negative (N) in the regression model.  
Since the true regression coefficients of the simulated data are known, we can calculate True Positive (TP), False Positive (FP), True Negative (TN), False Negative (FN), Recall (TP/(TP+FN)), and Precision (TP/(TP+FP)).

The weighted F-score is weighted on the Recall side by the importance $\theta$ as follows: 
\begin{equation}
\begin{aligned}
    F\theta \text{-score} &= 
    \frac{1 + \theta^{2}}{\frac{1}{\text{Precision}} + \frac{\theta^{2}}{\text{Recall}}} \\ 
    &= 
    \frac{(1 + \theta^{2})(\text{Precision} \times \text{Recall})}{\text{Recall} + \theta^{2} \times \text{Precision}}, 
    \nonumber
\end{aligned}
\end{equation}
where Precision = TP/(TP+FP) and Recall = TP/(TP+FN).
Since the screening needs to select as many important variables with non-zero regression coefficients as possible, we use the F2-score, which treats Recall as important.

\subsection{Simulation results}

\begin{figure}[htbp]
    \begin{tabular}{c c c}
      \begin{minipage}[t]{0.30\hsize}
        \centering
        $\varphi=0.5$
      \end{minipage} &
      \begin{minipage}[t]{0.30\hsize}
        \centering
        $\varphi=0.7$
      \end{minipage} &
      \begin{minipage}[t]{0.30\hsize}
        \centering
        $\varphi=0.9$
      \end{minipage} \\
      \begin{minipage}[t]{0.30\hsize}
        \centering
        \includegraphics[keepaspectratio, scale=0.50]{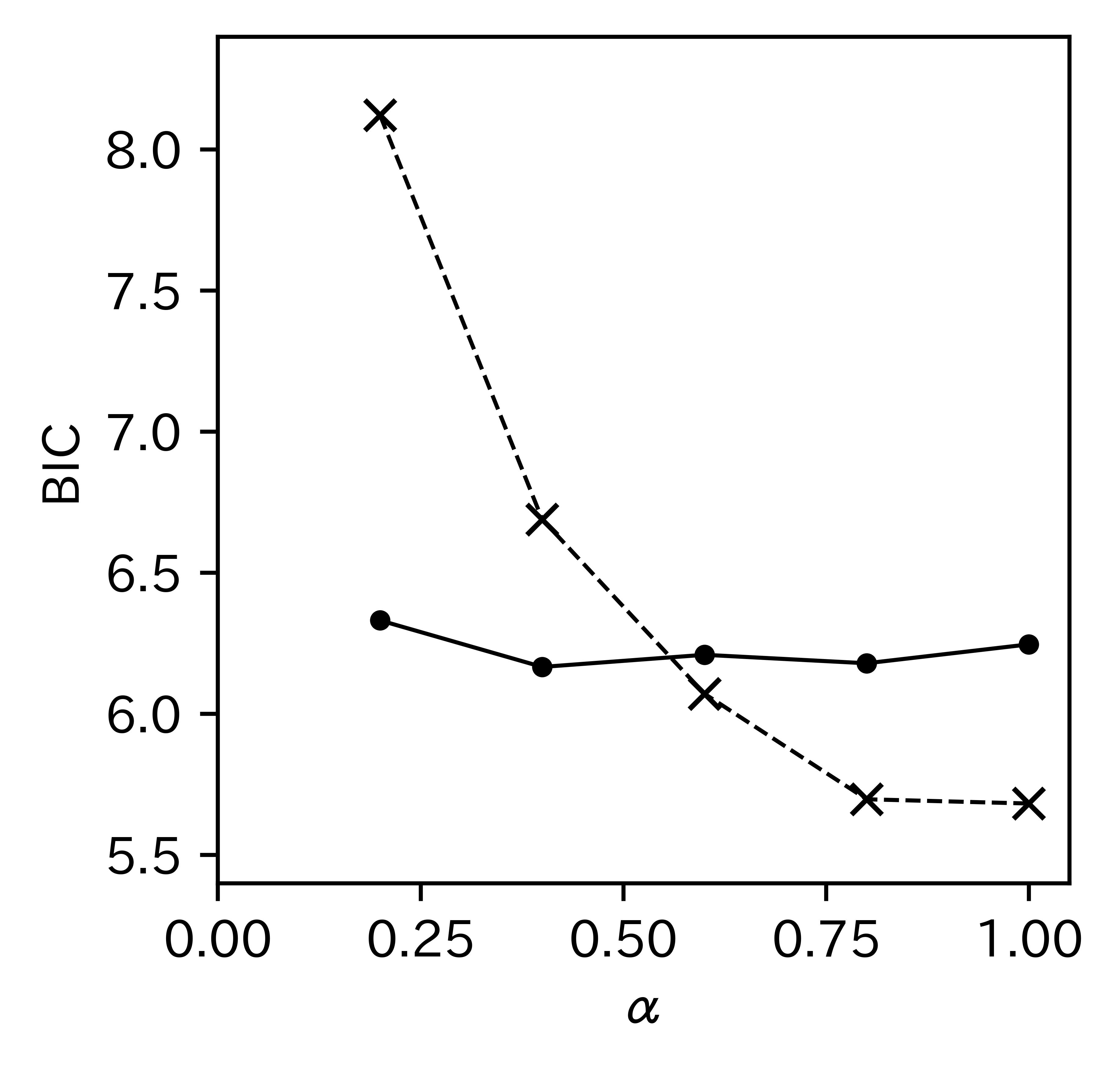}
        \label{example1_bic_phi05}
      \end{minipage} &
      \begin{minipage}[t]{0.30\hsize}
        \centering
        \includegraphics[keepaspectratio, scale=0.50]{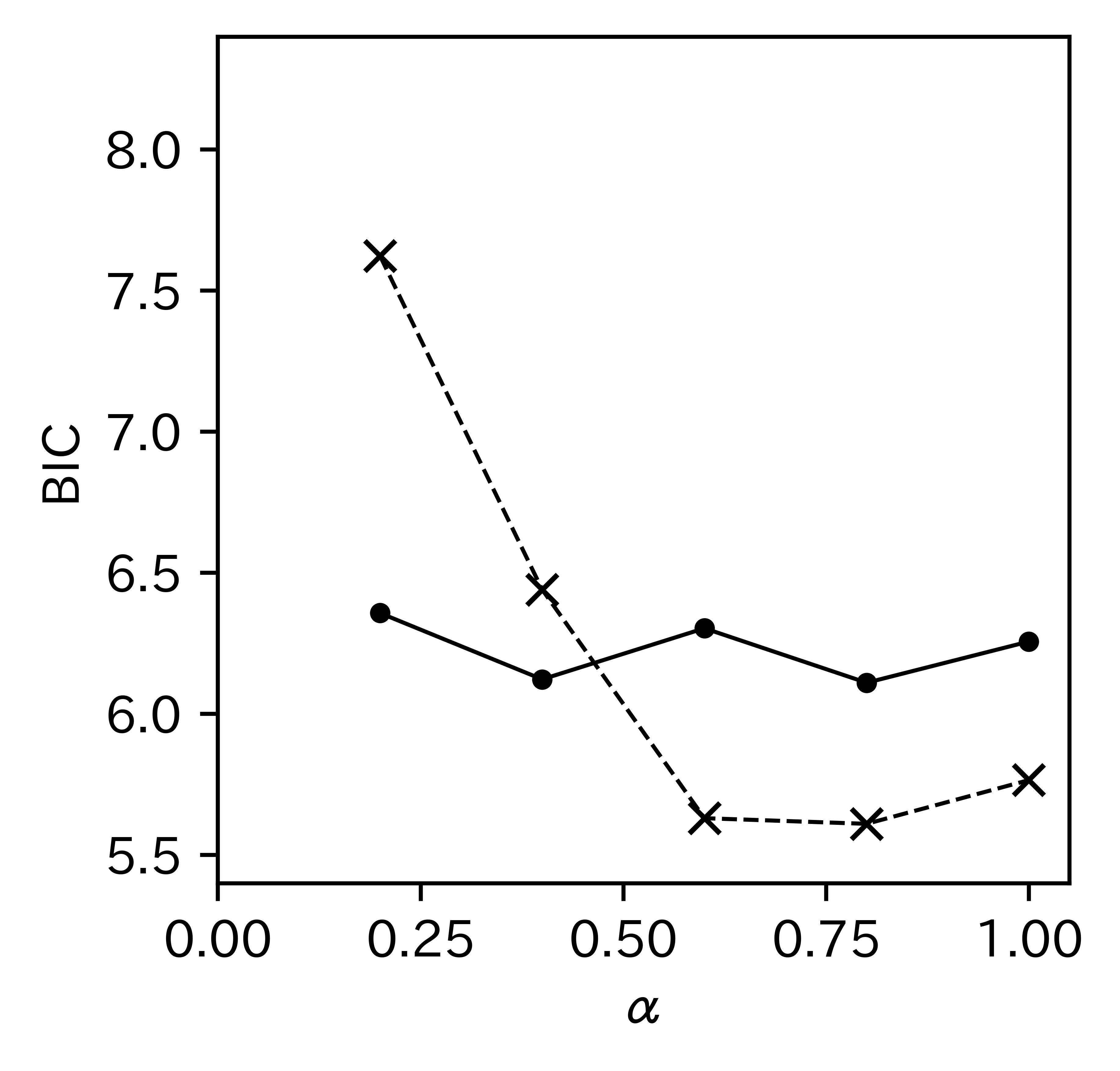}
        \label{example1_bic_phi07}
      \end{minipage} &
      \begin{minipage}[t]{0.30\hsize}
        \centering
        \includegraphics[keepaspectratio, scale=0.50]{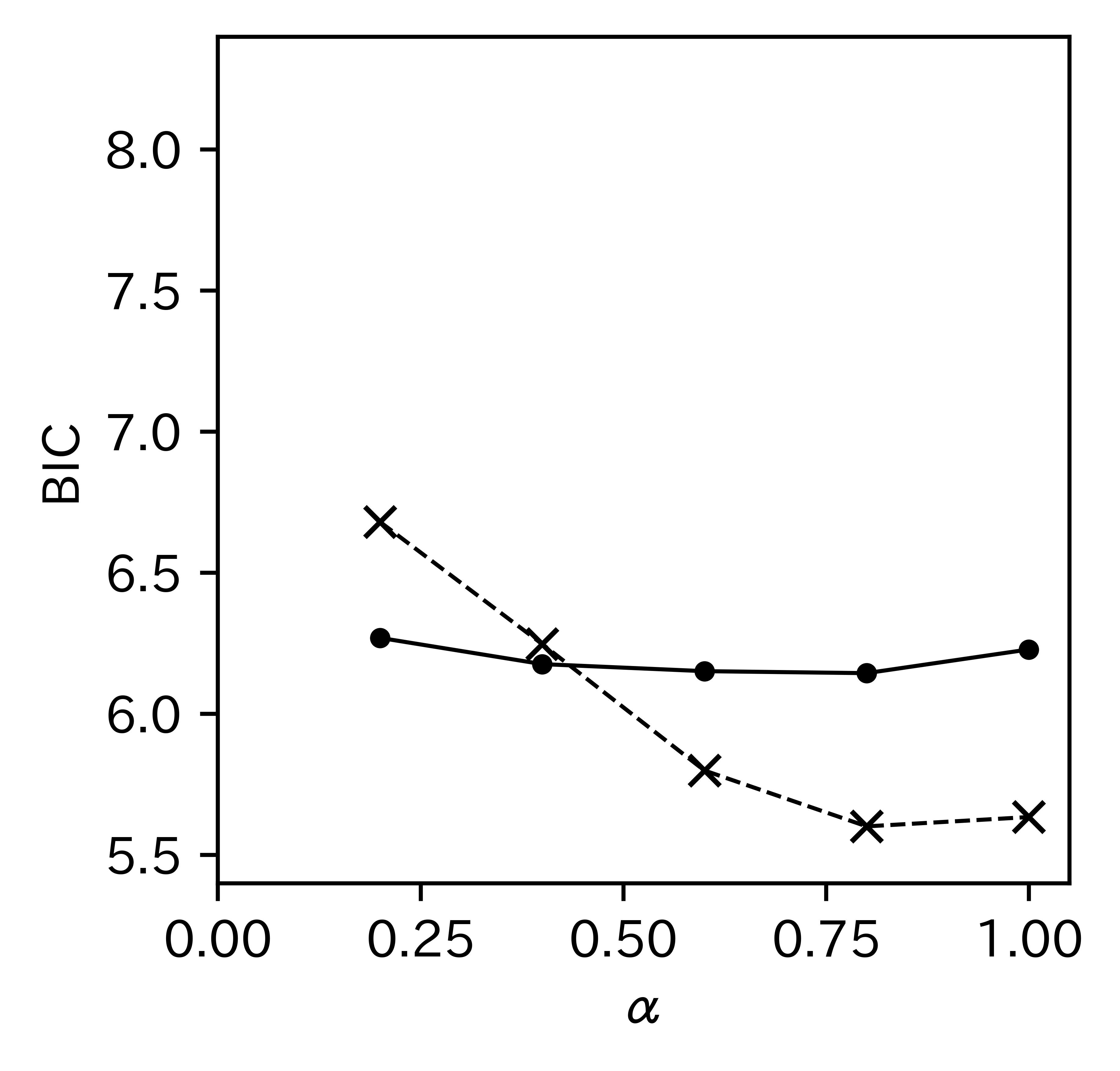}
        \label{example1_bic_phi09}
      \end{minipage} \\
      \begin{minipage}[t]{0.30\hsize}
        \centering
        \includegraphics[keepaspectratio, scale=0.50]{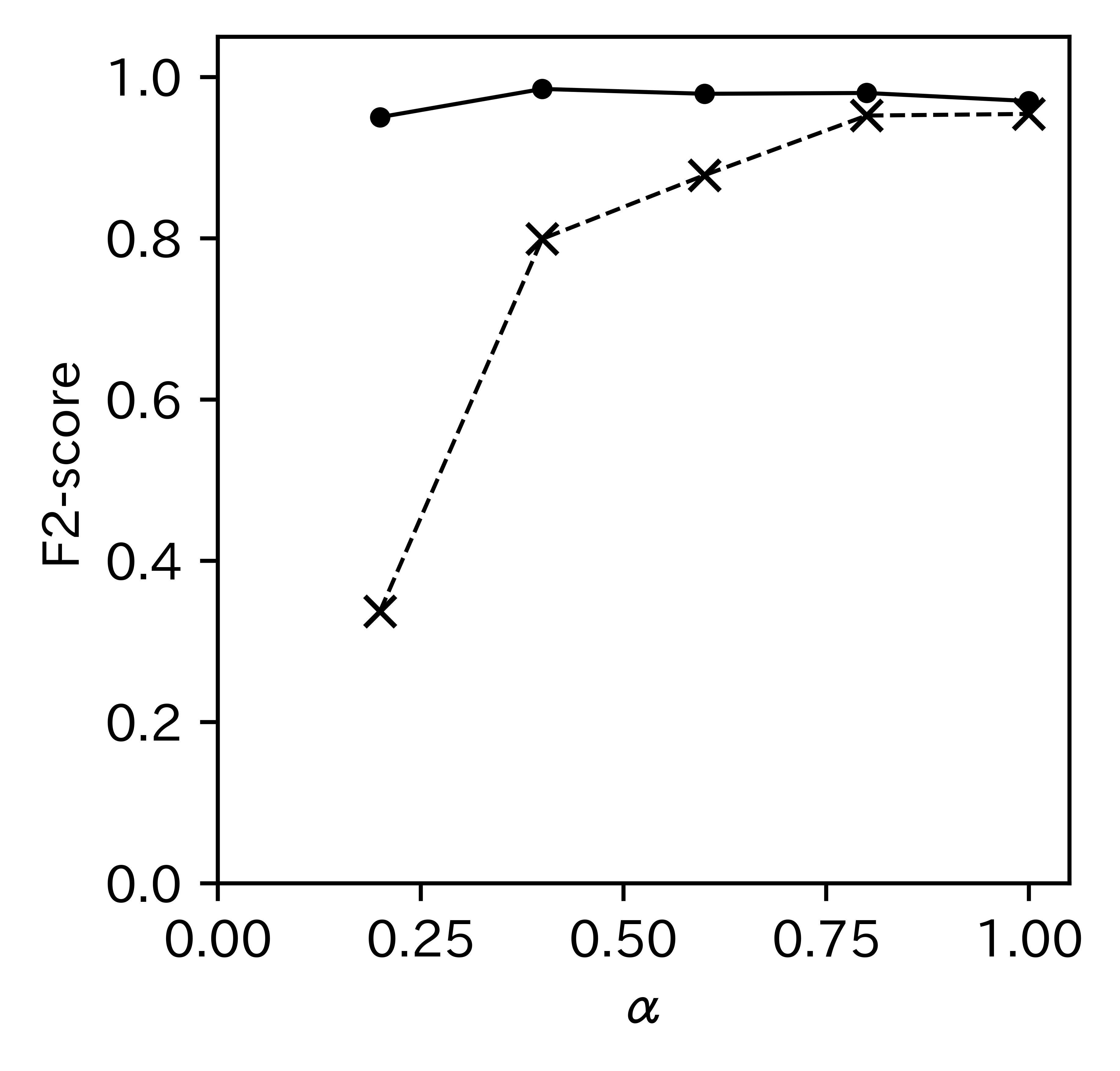}
        \label{example1_F2score_phi05}
      \end{minipage} &
      \begin{minipage}[t]{0.30\hsize}
        \centering
        \includegraphics[keepaspectratio, scale=0.50]{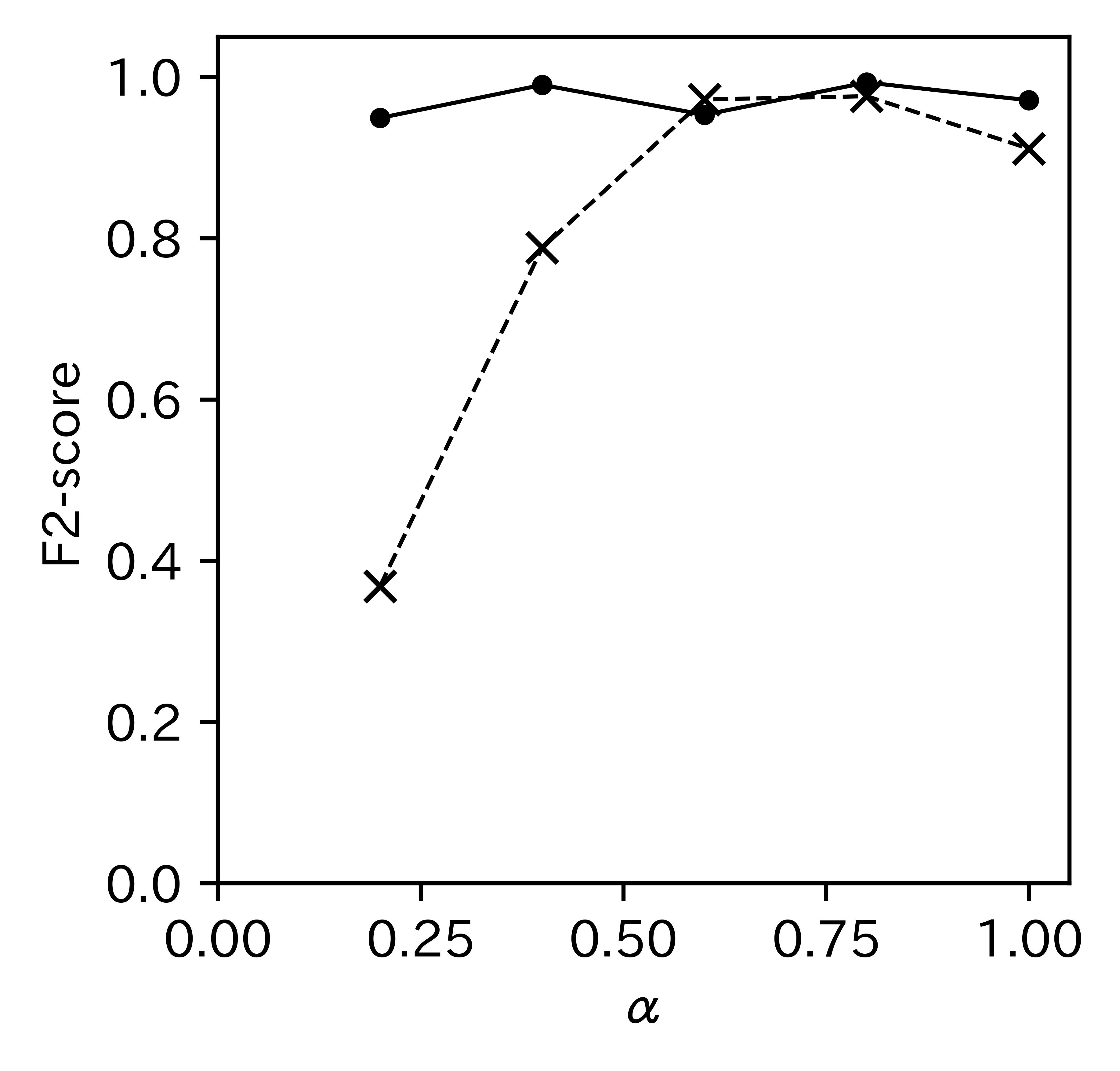}
        \label{example1_F2score_phi07}
      \end{minipage} &
      \begin{minipage}[t]{0.30\hsize}
        \centering
        \includegraphics[keepaspectratio, scale=0.50]{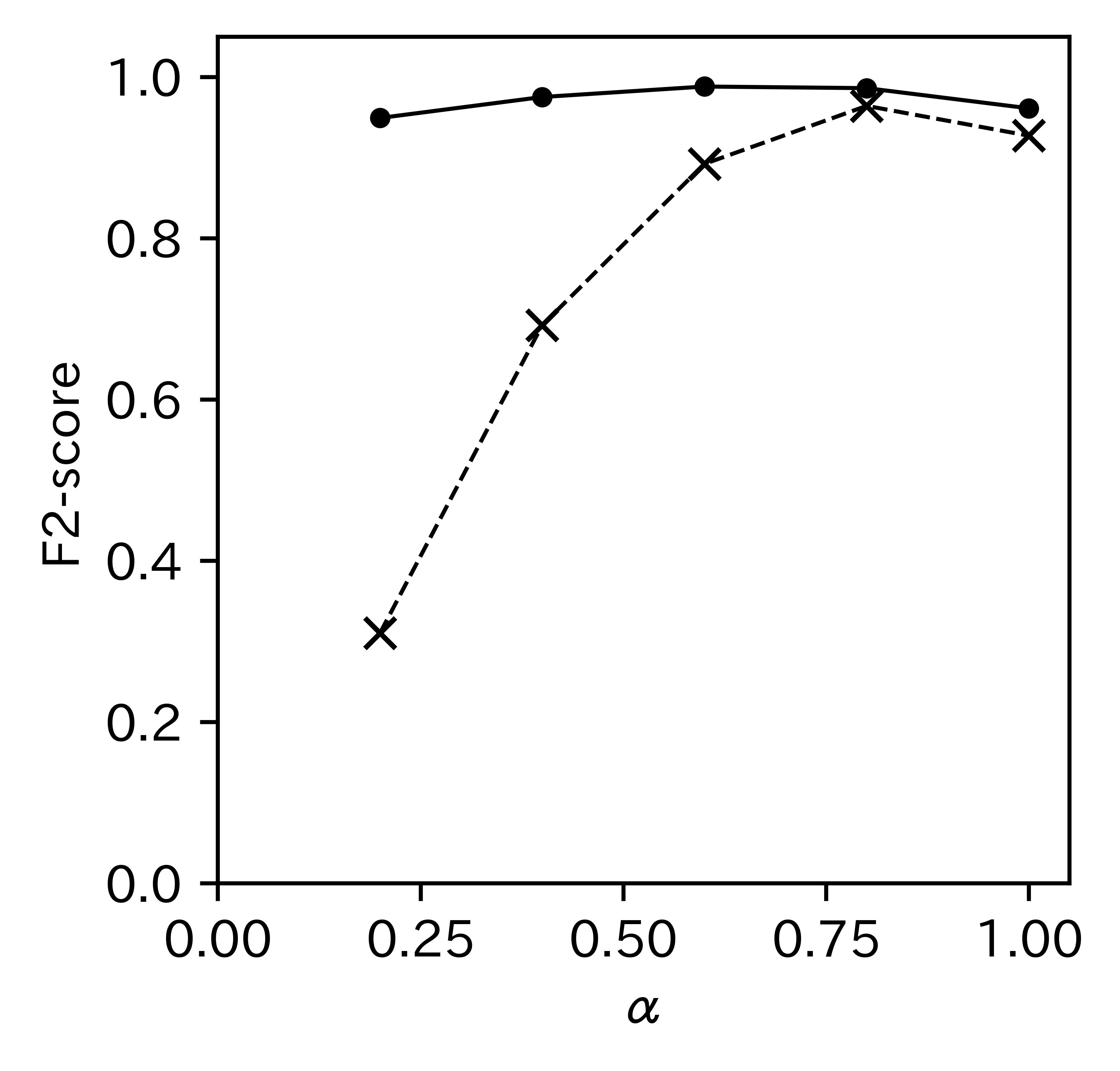}
        \label{example1_F2score_phi09}
      \end{minipage}
    \end{tabular}
\caption{
Values of BIC and F2-score for different $\alpha$ in TPPIS of Example 1.
The top row shows BIC results and the bottom row shows F2-score results.
The values for $n=300$ are represented by $\bullet$, and the values for $n=100$ are represented by $\times$.
$p$ is 1000 in all cases.
}
\label{fig:figure1}
\end{figure}

The results of the variable selection for Example 1 are shown in Table \ref{tb:example1}. 
The numbers in the $x_{(j)}$ column represent the total number of times that the $j$-th predictor variable is selected.
For all settings, SIS never selected $x_{(4)}$.
This is because $\bm{y}=(y_{1}, \ldots , y_{n})^{T}$ and $(x_{14}, \ldots , x_{n4})^{T}$ are uncorrelated due to the generation mechanism of the data, which gives a smaller $|\omega_{4}|$.
For other methods than SIS, the value of $|\omega_{4}|$ is larger than that for SIS due to the transformation process by factor analysis.
In particular, the proposed TPPIS obtains the largest $x_{(4)}$.
F2-scores for TPPIS are the highest under all settings.
Although the best $\alpha$ of TPPIS is 1 for the case $\varphi=0.5$, the F2-scores for TPPIS are better than those for PPIS because TPPIS selects $d$ by BIC.
We confirmed that the performance of TPPIS in variable selection is improved compared to the existing methods.
Figure \ref{fig:figure1} shows values of BIC and F2-scores for fixed $d$ and different $\alpha$ in TPPIS.
This figure demonstrates that $\alpha$ is selected appropriately by BIC.

The results for Example 2 are shown in Table \ref{tb:example2}.
The table shows that in many cases the numbers in $x_{(5)}$ are close to 100 because the fifth variable is uncorrelated with the other predictors.
F2-scores for TPPIS are the highest in all cases.

Table \ref{tb:example3} summarizes the result for Example 3. 
This shows that the numbers in $x_{(5)}$ and the value of F2-score are smaller than those of Example 2 due to the addition of the sixth variable, which is highly correlated with the fifth variable.
For the cases with $\varphi=0.9$, FPSIS$_{BIC}$ and TPPIS, which determine $d$ by BIC, give lower $x_{(6)}$ values.
It seems to be useful to use BIC to select $d$ for data with multicollinearity.
TPPIS gives the highest F2-score among all methods.

Table \ref{tb:example4-1} shows the results for Example 4.
In this example, the variables with large regression coefficients tend to be more important, resulting in $x_{(1)}\geq x_{(2)}\geq x_{(3)}\geq x_{(4)}$ under many settings.
F2-scores for PPIS and TPPIS are high because these methods are effective for the spike model.
In particular, TPPIS gives the highest F2-scores for all settings.

\section{Real data analysis}

We apply the proposed screening methods to the analysis of two real data sets.
For both datasets, we investigated TPPIS parameters $d$ and $\alpha$, as in Section \ref{Settings for simulated data}, and then the $d$ and $\alpha$ values giving the lowest BIC are selected as the optimal parameters.

\subsection{Condition monitoring of hydraulic systems}
\label{Condition monitoring of hydraulic systems}

We applied the screening methods to data on condition monitoring of a hydraulic system \cite{helwig2015condition}.
This dataset was obtained experimentally using a hydraulic test rig to measure values such as pressure, volumetric flow, and temperature while varying the settings of four different hydraulic components (coolers, valves, pumps, and accumulators).
We use data with the sample size 1449, taken under stable system settings.
The response is a value that expresses the degree of accumulator failure as a continuous value.
A higher value is closer to normal condition with 130 being the optimal pressure, 115 being a slightly reduced pressure, 100 being a severely reduced pressure, and 90 being close to total failure.
The predictors are the values measured by 17 sensors and form a total of 43680.
We apply the five screening methods to analyze this dataset as in the section on examples of simulated data.
The number of variables is determined using BIC.

Table \ref{tb:realdata1} shows the results of the analysis of this dataset.
From this result we find that TPPIS selects variables from the largest number of sensors.
TPPIS selects variables `volume flow sensors (FS)' and `efficiency factor (SE)', which are not selected by the other methods.
In addition, TPPIS gives the best BIC score among all methods.
These results indicate that these sensors may relate to the condition of accumulators.

\subsection{S\&P500}

The S\&P 500, one of the U.S. stock market indices, is obtained by weighting the market capitalization of 500 companies selected as representative of publicly traded companies.
This analysis uses the data for the year 2020.
The sample size is 253, which is the number of trading days.
The response is the value of the S\&P500, and the predictors are the stock price of each of the 500 companies that make up the S\&P500.
Note that the number of columns of predictors may be greater than 500 because some companies have multiple stocks, differentiated based on whether they include voting rights. 
Since the S\&P500 is weighted by market capitalization, it is assumed that the stock price of the company with the highest market capitalization is selected as an important variable.
The values of the S\&P500 are taken from FRED \cite{fred}, and the stock prices of the 500 companies that make up the S\&P500 are taken from 
\cite{500stocks}.

We applied five screening methods to this dataset and compared BIC and selected variables.
The results for the S\&P500 are shown in Table \ref{tb:realdata2}.
TPPIS gives the best BIC score among all the methods.
The 12 variables selected by TPPIS include companies with particularly large market capitalizations such as `AAPL' (Apple), `MSFT' (Microsoft) and `AMZN' (Amazon).

\section{Conclusion}

We have proposed TPPIS, a variable screening method for high-dimensional data with strong multicollinearity. 
TPPIS improves the variable selection performance by using a BIC-type criterion to determine the number of common factors that have a role in removing multicollinearity.
In the analysis of simulated data, TPPIS outperformed existing methods using factor analysis for variable selection. 
This suggests that TPPIS may be able to correctly select variables that are not considered important by existing methods.

The transformation process of TPPIS to remove multicollinearity from the data uses only information from the data corresponding to the predictors and we do not consider the relation to the response. 
Developing a transformation processing method that incorporates information from both types of data could further improve the variable selection performance. 
Although numerical examples confirmed that the performance of TPPIS is better than that of existing methods, no mathematical proof is provided. 
In the process of devising a proof, we may be able to identify the characteristics of the data for which TPPIS is most effective.

\section*{Acknowledgment}
This work was supported by JSPS KAKENHI Grant Numbers 19K11858 and 23K11005. 

\bibliographystyle{tfnlm}

\begin{thebibliography}{10}
	\providecommand{\url}[1]{\normalfont{#1}}
	\providecommand{\urlprefix}{Available from: }
	
	\bibitem{tian2015variable}
	Tian~S, Yu~Y, Guo~H. Variable selection and corporate bankruptcy forecasts. J
	Bank Financ. 2015;\hspace{0pt}52:89--100.
	
	\bibitem{fang2020predicting}
	Fang~T, Lee~TH, Su~Z. Predicting the long-term stock market volatility: A
	garch-midas model with variable selection. J Empir Finance.
	2020;\hspace{0pt}58:36--49.
	
	\bibitem{chowdhury2020variable}
	Chowdhury~MZI, Turin~TC. Variable selection strategies and its importance in
	clinical prediction modelling. Fam Med Community Health.
	2020;\hspace{0pt}8(1).
	
	\bibitem{yun2019overview}
	Yun~YH, Li~HD, Deng~BC, et~al. An overview of variable selection methods in
	multivariate analysis of near-infrared spectra. Trends Analyt Chem.
	2019;\hspace{0pt}113:102--115.
	
	\bibitem{fan2008sure}
	Fan~J, Lv~J. Sure independence screening for ultrahigh dimensional feature
	space. J R Stat Soc B. 2008;\hspace{0pt}70(5):849--911.
	
	\bibitem{fan2010sure}
	Fan~J, Song~R. Sure independence screening in generalized linear models with
	np-dimensionality. 2010;\hspace{0pt}.
	
	\bibitem{fan2011nonparametric}
	Fan~J, Feng~Y, Song~R. Nonparametric independence screening in sparse
	ultra-high-dimensional additive models. J Am Stat Assoc.
	2011;\hspace{0pt}106(494):544--557.
	
	\bibitem{li2012robust}
	Li~G, Peng~H, Zhang~J, et~al. Robust rank correlation based screening.
	2012;\hspace{0pt}.
	
	\bibitem{li2012feature}
	Li~R, Zhong~W, Zhu~L. Feature screening via distance correlation learning. J Am
	Stat Assoc. 2012;\hspace{0pt}107(499):1129--1139.
	
	\bibitem{balasubramanian2013ultrahigh}
	Balasubramanian~K, Sriperumbudur~B, Lebanon~G. Ultrahigh dimensional feature
	screening via rkhs embeddings. In: Proceedings of the Sixteenth International
	Conference on Artificial Intelligence and Statistics; 2013 29 April- 1 May;
	Scottsdale, Arizona. PMLR; 2013. p. 126--134.
	
	\bibitem{zhang2017correlation}
	Zhang~J, Liu~Y, Wu~Y. Correlation rank screening for ultrahigh-dimensional
	survival data. Comput Statist Data Anal. 2017;\hspace{0pt}108:121--132.
	
	\bibitem{fan2018sure}
	Fan~J, Lv~J. Sure independence screening. Wiley StatsRef: Stat Ref Online.
	2018;\hspace{0pt}.
	
	\bibitem{wang2016high}
	Wang~X, Leng~C. High dimensional ordinary least squares projection for
	screening variables. J R Stat Soc B. 2016;\hspace{0pt}:589--611.
	
	\bibitem{wang2012factor}
	Wang~H. Factor profiled sure independence screening. Biometrika.
	2012;\hspace{0pt}99(1):15--28.
	
	\bibitem{zhao2020high}
	Zhao~N, Xu~Q, Tang~ML, et~al. High-dimensional variable screening under
	multicollinearity. Stat. 2020;\hspace{0pt}9(1):e272.
	
	\bibitem{jia2012preconditioning}
	Jia~J, Rohe~K. Preconditioning to comply with the irrepresentable condition.
	arXiv preprint arXiv:12085584. 2012;\hspace{0pt}.
	
	\bibitem{helwig2015condition}
	Helwig~N, Pignanelli~E, Sch{\"u}tze~A. Condition monitoring of a complex
	hydraulic system using multivariate statistics. In: 2015 IEEE International
	Instrumentation and Measurement Technology Conference (I2MTC) Proceedings;
	2015 11-14 May; Pisa, Italy. IEEE; 2015. p. 210--215.
	
	\bibitem{fred}
	\text{FRED [Internet]. St. Louis: Federal Reserve Bank of St. Louis} ; 2023 Jan
	25 [cited 2023 Jan 25]. Available from:
	https://fred.stlouisfed.org/series/SP500.
	
	\bibitem{500stocks}
	Hanseo~P. Data from: \text{S\&P} 500 stocks price with financial statement
	[dataset] ; 2022 Apr 18 [cited 2023 Jan 25]. In: Kaggle Datasets [Internet].
	Available from:
	https://www.kaggle.com/hanseopark/sp-500-stocks-value-with-financial-statement.
	
\end{thebibliography}

\clearpage{}


\begin{table}[htbp]
\centering
\tbl{Simulation results for Example 1}
{\begin{tabular}{c c c c c c c c c c c}
\toprule
$n$     & $p$     & $\varphi$ & Method & best $\alpha$ & BIC & F2-score 
&$x_{(1)}$& $x_{(2)}$& $x_{(3)}$& $x_{(4)}$ \\ 
\midrule

100	&	1000	&	0.5	&	SIS	&	-	&	7.978	&	0.291	&	31	&	37	&	31	&	0	\\
	&		&		&	FPSIS	&	-	&	5.778	&	0.930	&	93	&	93	&	96	&	91	\\
	&		&		&	FPSIS$_{BIC}$	&	-	&	5.742	&	0.938	&	97	&	93	&	94	&	92	\\
	&		&		&	PPIS	&	-	&	5.682	&	0.954	&	96	&	95	&	97	&	94	\\
																					
	&		&		&	TPPIS	&	1.0	&	5.611	&	0.971	&	97	&	96	&	99	&	96	\\
	&		&	0.7	&	SIS	&	-	&	7.612	&	0.258	&	39	&	28	&	22	&	0	\\
	&		&		&	FPSIS	&	-	&	5.701	&	0.933	&	94	&	93	&	94	&	91	\\
	&		&		&	FPSIS$_{BIC}$	&	-	&	5.701	&	0.933	&	94	&	93	&	94	&	91	\\
	&		&		&	PPIS	&	-	&	5.765	&	0.911	&	93	&	92	&	91	&	88	\\
	&		&		&	TPPIS	&	0.8	&	5.610	&	0.976	&	99	&	97	&	98	&	98	\\
																					
	&		&	0.9	&	SIS	&	-	&	6.648	&	0.202	&	24	&	22	&	23	&	0	\\
	&		&		&	FPSIS	&	-	&	5.580	&	0.961	&	96	&	98	&	96	&	95	\\
	&		&		&	FPSIS$_{BIC}$	&	-	&	5.580	&	0.961	&	96	&	98	&	96	&	95	\\
	&		&		&	PPIS	&	-	&	5.634	&	0.927	&	91	&	95	&	94	&	90	\\
	&		&		&	TPPIS	&	0.8	&	5.601	&	0.964	&	96	&	98	&	98	&	96	\\

\midrule
 
300	&	1000	&	0.5	&	SIS	&	-	&	8.993	&	0.507	&	74	&	77	&	73	&	0	\\
	&		&		&	FPSIS	&	-	&	6.228	&	0.977	&	100	&	100	&	100	&	95	\\
	&		&		&	FPSIS$_{BIC}$	&	-	&	6.228	&	0.977	&	100	&	100	&	100	&	95	\\
	&		&		&	PPIS	&	-	&	6.246	&	0.970	&	99	&	99	&	98	&	95	\\
																					
	&		&		&	TPPIS	&	1.0	&	6.160	&	0.989	&	100	&	100	&	100	&	97	\\
	&		&	0.7	&	SIS	&	-	&	8.584	&	0.510	&	75	&	72	&	77	&	0	\\
	&		&		&	FPSIS	&	-	&	6.221	&	0.976	&	100	&	99	&	99	&	94	\\
	&		&		&	FPSIS$_{BIC}$	&	-	&	6.158	&	0.980	&	99	&	98	&	99	&	97	\\
	&		&		&	PPIS	&	-	&	6.256	&	0.971	&	98	&	100	&	98	&	93	\\
	&		&		&	TPPIS	&	0.8	&	6.110	&	0.993	&	100	&	100	&	100	&	99	\\
																					
	&		&	0.9	&	SIS	&	-	&	7.716	&	0.404	&	58	&	55	&	51	&	0	\\
	&		&		&	FPSIS	&	-	&	6.195	&	0.969	&	98	&	98	&	100	&	93	\\
	&		&		&	FPSIS$_{BIC}$	&	-	&	6.159	&	0.976	&	99	&	99	&	98	&	95	\\
	&		&		&	PPIS	&	-	&	6.228	&	0.961	&	99	&	98	&	99	&	92	\\
	&		&		&	TPPIS	&	0.8	&	6.144	&	0.986	&	100	&	100	&	100	&	97	\\
\bottomrule
\end{tabular}}
\label{tb:example1}
\end{table}


\begin{table}[htbp]
\centering
\tbl{Simulation results for Example 2}
{\begin{tabular}{c c c c c c c c c c c c c c}
\toprule
$n$     & $p$     & $\varphi$ & Method & best $\alpha$ & BIC & F2-score &
$x_{(1)}$& $x_{(2)}$& $x_{(3)}$& $x_{(4)}$& $x_{(5)}$ \\ 
\midrule
100	&	1000	&	0.5	&	SIS	&	-	&	8.236	&	0.264	&	4	&	5	&	3	&	0	&	100	\\
	&		&		&	FPSIS	&	-	&	6.336	&	0.921	&	93	&	93	&	94	&	93	&	99	\\
	&		&		&	FPSIS$_{BIC}$	&	-	&	6.336	&	0.921	&	93	&	93	&	94	&	93	&	99	\\
	&		&		&	PPIS	&	-	&	6.223	&	0.917	&	91	&	94	&	91	&	89	&	96	\\
																							
	&		&		&	TPPIS	&	1.0	&	6.183	&	0.931	&	94	&	94	&	94	&	92	&	98	\\
	&		&	0.7	&	SIS	&	-	&	7.726	&	0.245	&	0	&	2	&	1	&	0	&	100	\\
	&		&		&	FPSIS	&	-	&	6.821	&	0.858	&	84	&	81	&	85	&	91	&	99	\\
	&		&		&	FPSIS$_{BIC}$	&	-	&	6.481	&	0.905	&	90	&	89	&	94	&	97	&	100	\\
	&		&		&	PPIS	&	-	&	6.400	&	0.901	&	92	&	89	&	95	&	90	&	96	\\
																							
	&		&		&	TPPIS	&	1.0	&	6.211	&	0.934	&	95	&	93	&	96	&	95	&	99	\\
	&		&	0.9	&	SIS	&	-	&	6.737	&	0.238	&	0	&	0	&	0	&	0	&	100	\\
	&		&		&	FPSIS	&	-	&	7.840	&	0.389	&	21	&	21	&	28	&	86	&	35	\\
	&		&		&	FPSIS$_{BIC}$	&	-	&	6.182	&	0.893	&	83	&	88	&	88	&	91	&	97	\\
	&		&		&	PPIS	&	-	&	6.227	&	0.888	&	84	&	86	&	87	&	92	&	100	\\
																							
	&		&		&	TPPIS	&	1.0	&	6.126	&	0.918	&	89	&	89	&	94	&	91	&	98	\\
\midrule
300	&	1000	&	0.5	&	SIS	&	-	&	9.198	&	0.582	&	67	&	68	&	62	&	0	&	100	\\
	&		&		&	FPSIS	&	-	&	6.422	&	0.970	&	100	&	98	&	98	&	93	&	100	\\
	&		&		&	FPSIS$_{BIC}$	&	-	&	6.402	&	0.967	&	100	&	97	&	97	&	95	&	100	\\
	&		&		&	PPIS	&	-	&	6.452	&	0.975	&	100	&	100	&	100	&	92	&	100	\\
	&		&		&	TPPIS	&	0.6	&	6.255	&	0.990	&	100	&	100	&	100	&	99	&	100	\\
																							
	&		&	0.7	&	SIS	&	-	&	8.774	&	0.484	&	49	&	51	&	50	&	0	&	100	\\
	&		&		&	FPSIS	&	-	&	6.302	&	0.988	&	100	&	100	&	100	&	97	&	100	\\
	&		&		&	FPSIS$_{BIC}$	&	-	&	6.302	&	0.988	&	100	&	100	&	100	&	97	&	100	\\
	&		&		&	PPIS	&	-	&	6.280	&	0.990	&	100	&	100	&	100	&	98	&	100	\\
																							
	&		&		&	TPPIS	&	1.0	&	6.274	&	0.993	&	100	&	100	&	100	&	98	&	100	\\
	&		&	0.9	&	SIS	&	-	&	7.827	&	0.254	&	3	&	3	&	4	&	0	&	100	\\
	&		&		&	FPSIS	&	-	&	6.297	&	0.979	&	99	&	99	&	99	&	94	&	100	\\
	&		&		&	FPSIS$_{BIC}$	&	-	&	6.297	&	0.979	&	99	&	99	&	99	&	94	&	100	\\
	&		&		&	PPIS	&	-	&	6.314	&	0.970	&	98	&	97	&	98	&	94	&	100	\\
																							
	&		&		&	TPPIS	&	0.6	&	6.253	&	0.986	&	99	&	99	&	99	&	97	&	100	\\
\bottomrule
\end{tabular}}
\label{tb:example2}
\end{table}

\begin{table}[htbp]
\centering
\tbl{Simulation results for Example 3}
{\begin{tabular}{c c c c c c c c c c c c c}
\toprule
$n$     & $p$     & $\varphi$ & Method & best $\alpha$ & BIC & F2-score &
$x_{(1)}$& $x_{(2)}$& $x_{(3)}$& $x_{(4)}$& $x_{(5)}$& $x_{(6)}$ \\ 
\midrule
100	&	1000	&	0.5	&	SIS	&	-	&	8.215	&	0.000	&	0	&	0	&	0	&	0	&	0	&	100	\\
	&		&		&	FPSIS	&	-	&	6.724	&	0.798	&	86	&	85	&	86	&	88	&	80	&	99	\\
	&		&		&	FPSIS$_{BIC}$	&	-	&	6.609	&	0.826	&	94	&	89	&	88	&	90	&	77	&	99	\\
	&		&		&	PPIS	&	-	&	6.273	&	0.907	&	97	&	95	&	96	&	96	&	91	&	99	\\
	&		&		&	TPPIS	&	1.0	&	6.273	&	0.907	&	97	&	95	&	96	&	96	&	91	&	99	\\
																									
	&		&	0.7	&	SIS	&	-	&	7.737	&	0.000	&	0	&	0	&	0	&	0	&	0	&	100	\\
	&		&		&	FPSIS	&	-	&	6.555	&	0.780	&	85	&	84	&	85	&	85	&	69	&	100	\\
	&		&		&	FPSIS$_{BIC}$	&	-	&	6.468	&	0.837	&	91	&	89	&	91	&	93	&	76	&	100	\\
	&		&		&	PPIS	&	-	&	6.470	&	0.838	&	91	&	89	&	90	&	92	&	81	&	100	\\
																									
	&		&		&	TPPIS	&	1.0	&	6.274	&	0.887	&	95	&	94	&	95	&	96	&	82	&	99	\\
	&		&	0.9	&	SIS	&	-	&	6.773	&	0.000	&	0	&	0	&	0	&	0	&	0	&	100	\\
	&		&		&	FPSIS	&	-	&	7.220	&	0.401	&	37	&	31	&	33	&	93	&	1	&	97	\\
	&		&		&	FPSIS$_{BIC}$	&	-	&	6.364	&	0.831	&	87	&	85	&	87	&	92	&	76	&	66	\\
	&		&		&	PPIS	&	-	&	6.346	&	0.823	&	88	&	88	&	86	&	90	&	79	&	100	\\
																									
	&		&		&	TPPIS	&	1.0	&	6.282	&	0.879	&	92	&	92	&	93	&	95	&	83	&	64	\\
\midrule

300	&	1000	&	0.5	&	SIS	&	-	&	9.290	&	0.382	&	45	&	52	&	52	&	0	&	62	&	100	\\
	&		&		&	FPSIS	&	-	&	6.510	&	0.923	&	97	&	97	&	97	&	94	&	95	&	100	\\
	&		&		&	FPSIS$_{BIC}$	&	-	&	6.460	&	0.927	&	97	&	97	&	97	&	96	&	97	&	100	\\
	&		&		&	PPIS	&	-	&	6.443	&	0.941	&	98	&	99	&	99	&	96	&	98	&	100	\\
																									
	&		&		&	TPPIS	&	1.0	&	6.396	&	0.954	&	100	&	100	&	100	&	98	&	100	&	100	\\
	&		&	0.7	&	SIS	&	-	&	8.845	&	0.217	&	32	&	32	&	31	&	0	&	33	&	100	\\
	&		&		&	FPSIS	&	-	&	6.477	&	0.925	&	97	&	97	&	98	&	95	&	97	&	100	\\
	&		&		&	FPSIS$_{BIC}$	&	-	&	6.415	&	0.944	&	99	&	99	&	99	&	97	&	99	&	99	\\
	&		&		&	PPIS	&	-	&	6.379	&	0.950	&	99	&	99	&	100	&	98	&	99	&	100	\\
	&		&		&	TPPIS	&	1.0	&	6.379	&	0.950	&	99	&	99	&	100	&	98	&	99	&	100	\\
																									
	&		&	0.9	&	SIS	&	-	&	7.883	&	0.006	&	1	&	0	&	1	&	0	&	1	&	100	\\
	&		&		&	FPSIS	&	-	&	6.610	&	0.871	&	95	&	95	&	95	&	92	&	95	&	100	\\
	&		&		&	FPSIS$_{BIC}$	&	-	&	6.398	&	0.937	&	94	&	94	&	95	&	91	&	100	&	22	\\
	&		&		&	PPIS	&	-	&	6.477	&	0.882	&	92	&	92	&	92	&	91	&	92	&	100	\\
																									
	&		&		&	TPPIS	&	0.8	&	6.305	&	0.975	&	99	&	99	&	99	&	97	&	100	&	27	\\
\bottomrule        
\end{tabular}}
\label{tb:example3}
\end{table}

\begin{table}[htbp]
\centering
\tbl{Simulation results for Example 4}
{\begin{tabular}{c c c c c c c c c c c c c c}
\toprule
$n$     & $p$     & $d$ & $m$ & Method & best $\alpha$ & BIC & F2-score & $x_{(1)}$& $x_{(2)}$& $x_{(3)}$& $x_{(4)}$ \\ 
\midrule
100	&	1000	&	3	&	20	&	SIS	&	-	&	10.595	&	0.332	&	100	&	5	&	11	&	0	\\
	&		&		&		&	FPSIS	&	-	&	9.732	&	0.779	&	100	&	100	&	95	&	3	\\
	&		&		&		&	FPSIS$_{BIC}$	&	-	&	9.530	&	0.818	&	100	&	100	&	90	&	36	\\
	&		&		&		&	PPIS	&	-	&	9.432	&	0.836	&	100	&	100	&	95	&	33	\\
																							
	&		&		&		&	TPPIS	&	1.0	&	9.413	&	0.839	&	100	&	100	&	92	&	40	\\
	&		&		&	40	&	SIS	&	-	&	10.946	&	0.441	&	100	&	29	&	35	&	0	\\
	&		&		&		&	FPSIS	&	-	&	10.329	&	0.771	&	100	&	100	&	96	&	0	\\
	&		&		&		&	FPSIS$_{BIC}$	&	-	&	10.122	&	0.800	&	100	&	98	&	88	&	34	\\
	&		&		&		&	PPIS	&	-	&	9.963	&	0.843	&	100	&	100	&	97	&	39	\\
	&		&		&		&	TPPIS	&	1.0	&	9.963	&	0.843	&	100	&	100	&	97	&	39	\\
																							
	&		&		&	60	&	SIS	&	-	&	11.098	&	0.588	&	100	&	65	&	61	&	0	\\
	&		&		&		&	FPSIS	&	-	&	10.808	&	0.743	&	100	&	99	&	85	&	0	\\
	&		&		&		&	FPSIS$_{BIC}$	&	-	&	10.808	&	0.743	&	100	&	99	&	85	&	0	\\
	&		&		&		&	PPIS	&	-	&	10.563	&	0.815	&	100	&	100	&	96	&	26	\\
	&		&		&		&	TPPIS	&	1.0	&	10.563	&	0.815	&	100	&	100	&	96	&	26	\\
																							
	&		&		&	80	&	SIS	&	-	&	11.347	&	0.679	&	100	&	80	&	84	&	0	\\
	&		&		&		&	FPSIS	&	-	&	11.158	&	0.757	&	100	&	99	&	91	&	0	\\
	&		&		&		&	FPSIS$_{BIC}$	&	-	&	11.158	&	0.757	&	100	&	99	&	91	&	0	\\
	&		&		&		&	PPIS	&	-	&	11.031	&	0.794	&	100	&	100	&	95	&	11	\\
	&		&		&		&	TPPIS	&	1.0	&	11.031	&	0.794	&	100	&	100	&	95	&	11	\\
\midrule
300	&	1000	&	3	&	60	&	SIS	&	-	&	11.435	&	0.716	&	100	&	96	&	99	&	0	\\
	&		&		&		&	FPSIS	&	-	&	11.253	&	0.778	&	100	&	100	&	100	&	0	\\
	&		&		&		&	FPSIS$_{BIC}$	&	-	&	10.334	&	0.970	&	100	&	100	&	100	&	95	\\
	&		&		&		&	PPIS	&	-	&	10.249	&	0.994	&	100	&	100	&	100	&	99	\\
	&		&		&		&	TPPIS	&	1.0	&	10.249	&	0.994	&	100	&	100	&	100	&	99	\\
																							
	&		&		&	120	&	SIS	&	-	&	12.240	&	0.773	&	100	&	100	&	100	&	0	\\
	&		&		&		&	FPSIS	&	-	&	12.223	&	0.784	&	100	&	100	&	100	&	0	\\
	&		&		&		&	FPSIS$_{BIC}$	&	-	&	11.757	&	0.927	&	100	&	100	&	100	&	77	\\
	&		&		&		&	PPIS	&	-	&	11.688	&	0.948	&	100	&	100	&	100	&	85	\\
	&		&		&		&	TPPIS	&	1.0	&	11.688	&	0.948	&	100	&	100	&	100	&	85	\\
																							
	&		&		&	180	&	SIS	&	-	&	12.928	&	0.784	&	100	&	100	&	100	&	0	\\
	&		&		&		&	FPSIS	&	-	&	12.918	&	0.788	&	100	&	100	&	100	&	0	\\
	&		&		&		&	FPSIS$_{BIC}$	&	-	&	12.804	&	0.845	&	100	&	100	&	94	&	45	\\
	&		&		&		&	PPIS	&	-	&	12.702	&	0.888	&	100	&	100	&	99	&	57	\\
	&		&		&		&	TPPIS	&	1.0	&	12.702	&	0.888	&	100	&	100	&	99	&	57	\\
																							
	&		&		&	240	&	SIS	&	-	&	13.448	&	0.787	&	100	&	100	&	100	&	0	\\
	&		&		&		&	FPSIS	&	-	&	13.441	&	0.789	&	100	&	100	&	100	&	0	\\
	&		&		&		&	FPSIS$_{BIC}$	&	-	&	13.441	&	0.789	&	100	&	100	&	100	&	0	\\
	&		&		&		&	PPIS	&	-	&	13.416	&	0.815	&	100	&	100	&	92	&	27	\\
	&		&		&		&	TPPIS	&	0.8	&	13.408	&	0.823	&	100	&	100	&	94	&	30	\\
\bottomrule
\end{tabular}}
\label{tb:example4-1}
\end{table}

\begin{table}[htbp]
\centering
\tbl{Results for condition monitoring of hydraulic systems}
{\begin{tabular}{c c c c }
\toprule
Method & best $\alpha$ & BIC    & Number of selected variables\\
\midrule
SIS     &	-	&	6.300	&	12	\\
FPSIS	&	-	&	7.278	&	1	\\
FPSIS$_{BIC}$	&	-	&	7.027	&	14	\\
PPIS	&	-	&	7.051	&	41	\\
TPPIS	&	0.4	&	6.271	&	17	\\
\bottomrule
\end{tabular}}
\label{tb:realdata1}
\end{table}
\begin{table}[htbp]
\centering
\tbl{Results for S\&P500}
{\begin{tabular}{c c c c }
\toprule
Method & best $\alpha$ & BIC    & Number of selected variables\\ \midrule
SIS	    &	-	&	2.118	&	2\\
FPSIS	&	-	&	1.986	&	5\\
FPSIS$_{BIC}$	&	-	&	1.387	&	12\\
PPIS	&	-	&	1.818	&	11\\
TPPIS	&	0.4	&	1.101	&	7\\
\bottomrule
\end{tabular}}
\label{tb:realdata2}
\end{table}

\end{document}